\documentclass[aps, prd,preprint,showpacs,superscriptaddress,groupedaddress,nofootinbib]{revtex4-1}
\usepackage{amsmath,amsfonts}
\usepackage[dvips]{graphicx}
\usepackage{morefloats}
\graphicspath{{Fig/}}
\begin{document}

\title{On the viability of the truncated Israel-Stewart theory \\ in cosmology}
\author{Dmitry Shogin}
\email{dmitry.shogin@uis.no}
\affiliation{Faculty of Science and Technology, University of Stavanger, N-4036 Stavanger, Norway}
\author{Per Amund Amundsen}
\email{per.a.amundsen@uis.no}
\affiliation{Faculty of Science and Technology, University of Stavanger, N-4036 Stavanger, Norway}
\author{Sigbj\o rn Hervik}
\email{sigbjorn.hervik@uis.no}
\affiliation{Faculty of Science and Technology, University of Stavanger, N-4036 Stavanger, Norway}

\begin{abstract}
We apply the causal Israel-Stewart theory of irreversible thermodynamics to model the matter content of the universe as a dissipative fluid with bulk and shear viscosity. Along with the full transport equations we consider their widely used truncated version. By implementing a dynamical systems approach to Bianchi type~IV and~V cosmological models with and without cosmological constant, we determine the future asymptotic states of such universes and show that the truncated Israel-Stewart theory leads to solutions essentially different from the full theory. The solutions of the truncated theory may also manifest unphysical properties. Finally, we find that in the full theory shear viscosity can give a substantial rise to dissipative fluxes, driving the fluid extremely far from equilibrium, where the linear Israel-Stewart theory ceases to be valid.
\end{abstract}


\maketitle

\section{Introduction}
\label{Sec:Intro}
In relativistic cosmology, the energy-matter content of the universe is often considered to be a perfect fluid in equilibrium. Perfect fluids manifest no dissipative effects and do not generate entropy, their dynamics being reversible. The early universe is commonly modelled by a radiation fluid, while for the later epoch a dust model is usually applied. However, the forces driving the fluid towards the equilibrium state will always be dissipative. Also, the transition between the epochs involves interactions between radiation and matter, which again implies the presence of dissipative processes~\cite{Udey1982,Zimdahl1996,Zimdahl1997}. Such processes are described in the framework of dissipative, irreversible thermo- and fluiddynamics.
\par 
The potential importance of dissipative processes in cosmology was briefly discussed by R.~Tolman already in 1934~\cite{Tolman1934}. The relativistic formulation of fluid dynamics goes back to Eckart~\cite{Eckart1940} and Landau and Lifshitz~\cite{Landau1958}, who pointed out that all real substances, which should include cosmic fluids, do have transport properties such as viscosity and heat conductivity; for applications to cosmology, see e.\,g.~\cite{Kolb1990}. However, it is now well known that both variants ot the theory suffer from non-causality, including dissipative perturbations that propagate at infinite speeds, and instabilities~\cite{Hiscock1983, Hiscock1985, Hiscock1988a}; such features are clearly not acceptable in cosmology.
\par 
These pathologies have been mediated in the more advanced Israel-Stewart~(IS) theory of irreversible thermodynamics~\cite{Israel1976, Israel1979}; for a brief review with applications to cosmology, see~\cite{Maartens1996}. Stability and causality are provided by including terms up to second order in the dissipative fluxes in the 4-vector of entropy production, while in the Eckart theory and the Landau-Lifshitz formulation, these terms are not included.
\par 
A simplified version of the IS~transport equations is often employed instead of the full theory~\cite{Maartens1996}. In this truncated theory, several divergence terms are assumed to be negligible and are omitted, without violating causality and stability.
\par 
Standard Friedmann-Robertson-Walker cosmologies have been studied using both truncated~\cite{Coley1995, Zimdahl1997a} and full~\cite{Maartens1995, Coley1996} versions of the IS~theory. In particular, reheating and inflation within causal thermodynamics have been considered respectively in~\cite{Zimdahl1997a} and~\cite{Maartens1995}. It was reported that some solutions obtained using the truncated theory are substantially different from those of the full theory, while the others are qualitatively similar.
\par 
The geometry of spatially homogeneous and isotropic cosmological models restricts the dissipative processes to so-called scalar dissipation, i.\thinspace e. shear viscosity and heat conduction are not allowed in this case. The next step in investigating dissipative cosmologies therefore involves anisotropic models. Spatially homogeneous Bianchi cosmologies~\cite{Ellis2012, Groen2007, Christiansen2008} are of great theoretical interest and have been subject to thorough studies~(see e.\thinspace g.~\cite{Ellis2012} and references therein; for causal thermodynamics applied to simple anisotropic models, see~\cite{Belinskii1979, Hoogen1995}).
\par 
In this paper we shall investigate the application of the IS theory to important Bianchi class~B cosmological models, namely types~IV and~V, and find out to which extent the truncated version of the IS equations represents the results of the full theory in anisotropic spacetimes. We shall use a simplified fluid model, which, however, retains the basic transport properties.
\par
The paper is organized as follows. In section~\ref{Sec:Model} we briefly describe the dissipative fluid model and the assumptions made. The system of equations governing the dynamics of the cosmologies is presented in section~\ref{Sec:Equations}. The future asymptotic states of the models without and with a positive cosmological constant are described in sections~\ref{Sec:AttractorsNoLambda} and~\ref{Sec:AttractorsLambda}, respectively. Conclusions are discussed in section~\ref{Sec:Conslusion}.

\section{The fluid model}
\label{Sec:Model}
We consider the matter content of the universe to be a dissipative fluid with energy-momentum tensor 
\begin{equation}
\label{Eq:Model:Tensor}
T_{\alpha \beta} = (\rho+p+\pi)u_\alpha u_\beta+(p+\pi)g_{\alpha \beta}+q_{(\alpha}u_{\beta)}+\pi_{\alpha \beta},
\end{equation}
where~$\pi$ stands for the bulk viscous pressure, and~$\pi_{\alpha\beta}$ is the shear viscous stress tensor (with~$\pi_{\alpha\beta}u^\beta=\pi_{[\alpha\beta]}=\pi_\alpha^{~\alpha}=0$), the rest of the notations being standard. Pressure~$p$ and energy density~$\rho$ in~(\ref{Eq:Model:Tensor}) refer to the local equilibrium values of the corresponding functions.
\par 
The energy-momentum tensor is decomposed with respect to a unit vector~$\mathbf{u}$. We apply the orthonormal frame approach~\cite{Elst1997}, where~$\mathbf{u}$ is chosen to be a unit vector, normal to spatial hypersurfaces. In the current paper we shall treat the case of so-called non-tilted fluids~\cite{King1973}, the 4-velocity of which, as defined by the particle flow, is aligned with~$\mathbf{u}$.

\par 
For simplicity, we consider the so-called $\gamma$-fluids with barotropic pressure and temperature obeying
\begin{equation}
p=(\gamma-1)\rho, \quad T\propto \rho^{(\gamma-1)/\gamma}, \label{Eq:Model:Gamma-Law}
\end{equation}
the~$\gamma$-parameter being a positive constant. We assume~$1<\gamma \leq 4/3$, which is often used to model a dust-radiation mixture as a single fluid.

\par 
We take into account that the linear transport equations of the IS~theory are derived with the assumption that the fluid is close to equilibrium. This implies that the dissipative fluxes are small and satisfy
\begin{equation}
\vert \pi \vert <<p, \qquad {(\pi_{ab}\pi^{ab})}^{1/2}<<p, \qquad {(q_aq^a)}^{1/2}<<\rho. \label{Eq:Model:EqulibriumConditions}
\end{equation}

Heat conductivity of the fluid is known to cause problems in the linear Israel-Stewart theory, since corresponding deviations from equilibrium~(\ref{Eq:Model:EqulibriumConditions}) may become infinitely large~\cite{Hiscock1988}. We therefore assume that the fluid is not heat-conducting, which in a non-tilted cosmological model leads to a vanishing energy flux vector $(q_a=0)$.

\par 
In certain solutions the near-equilibrium conditions~(\ref{Eq:Model:EqulibriumConditions}) are broken by bulk or shear viscous stresses. For example, bulk viscous terms are large if the fluid undergoes accelerated expansion. Strictly speaking, a consistent theory of non-linear thermodynamics is required to provide an accurate description of out-of-equilibrium processes~\cite{Maartens1997,Chimento1997}. We nevertheless assume that in case of finite and relatively small deviations from equilibrium, the solutions obtained using linear thermodynamics are still physically reasonable.

The transport equations of IS~theory in their full form can be written as\footnote{Index 1 is reserved for heat conduction, which is neglected in the present paper}
\begin{align}
\tau_0\dot{\pi}+\pi &= -3\zeta H-\frac{1}{2}\tau_0\pi \left[3H+\frac{\dot{\tau}_0}{\tau_0}-\frac{\dot{\zeta}}{\zeta}-\frac{\dot{T}}{T} \right],\label{Eq:Model:IS-Full-0-Bulk} \\
\tau_2\dot{\pi}_{ab}+\pi_{ab} &= -2\eta \sigma_{ab}-\frac{1}{2}\tau_2\pi_{ab} \left[3H+\frac{\dot{\tau}_2}{\tau_2}-\frac{\dot{\eta}}{\eta}-\frac{\dot{T}}{T} \right], \label{Eq:Model:IS-Full-2-Shear}
\end{align}
where the (positive) bulk and shear viscosity coefficients $\zeta$ and $\eta$ are related to the corresponding relaxation times $\tau_0, \tau_2$ by
\begin{equation}
\tau_0=\zeta \beta_0, \qquad \tau_2=2\eta \beta_2, \label{Eq:Model:RelaxationTimes}
\end{equation}
$\beta_0, \beta_2$ being the non-negative thermodynamic coefficients for scalar and tensor dissipative contributions to the entropy density.
\par
The terms in square brackets on the right-hand sides of~(\ref{Eq:Model:IS-Full-0-Bulk})--(\ref{Eq:Model:IS-Full-2-Shear}) are often implicitly supposed to be negligible in comparison with the other terms in the equations. This leads to a widely used simplified version of equations~(\ref{Eq:Model:IS-Full-0-Bulk})--(\ref{Eq:Model:IS-Full-2-Shear}), which is commonly referred to as the truncated Israel-Stewart theory~\cite{Maartens1995}.

\par 
For simplicity, we make the widely used assumption that the transport coefficients and corresponding relaxation times~(\ref{Eq:Model:RelaxationTimes}) are proportional to powers of the energy density of the fluid:
\begin{align}
\label{Eq:Model:ViscIndices}
\begin{split}
\zeta & \propto \rho^{m_0}, \qquad \frac{1}{\beta_0} \propto \rho^{r_0}; \\
\eta  & \propto \rho^{m_2}, \qquad \frac{1}{\beta_2} \propto \rho^{r_2}.
\end{split}
\end{align}
There are some discussions in literature concerning what the exponents should actually be, see e.\thinspace g. \cite{Coley1995, Hoogen1995}. However, the only choice that results in values consistent with correct dimensional scaling properties is:
\begin{equation}
\label{Eq:Model:PhysicalIndices}
m_0=m_2=1/2, \qquad r_0=r_2=1.
\end{equation}

Finally, we note an important point. The hydrodynamic description must be valid for the considered matter model, i.\thinspace e. the mean interaction time~$t_c$ of fluid particles is required to be much shorter than the characteristic timescale for macroscopic processes. In cosmology, this means that
\begin{equation}
t_c<<H^{-1}. \label{Eq:Model:HydrodynamicDescription}
\end{equation}
Therefore, we assume the existence of some long-range interactions, which ensure that the fluid description~(\ref{Eq:Model:HydrodynamicDescription}) remains valid for the matter under accelerated expansion.

\section{Evolution equations and constraints}
\label{Sec:Equations}

We investigate the chosen cosmological models using the dynamical systems approach~\cite{Wainwright1997}. Scale-invariant dimensionless geometric and fluid variables are introduced using the common principles and notations of the field. Dimensionless bulk and shear viscosity are introduced by
\begin{align}
\frac{\pi}{3H^2} &= \Pi; \\
\frac{\pi_{ab}}{H^2} &= \Pi_{ab}= \left[ 
\begin{array}{ccc}
-2\Pi_+ & 0 & 0 \\
0 & \Pi_+ +\sqrt{3}\Pi_- & \sqrt{3}\Pi_{23} \\
0 & \sqrt{3}\Pi_{23} & \Pi_+-\sqrt{3}\Pi_-
\end{array} \right],
\end{align}
where we have taken into account that~$\Pi_{12}=\Pi_{13}=\Sigma_{12}=\Sigma_{13}=0$ for non-tilted cosmologies. With the choice~(\ref{Eq:Model:PhysicalIndices}) we can introduce positive constants~$a_0,~b_0$ and~$a_2,~b_2$ -- the bulk and shear viscosity parameters, respectively -- by
\begin{align}
\begin{split}
\frac{1}{H^2}\cdot\frac{1}{\beta_0} &= a_0 \Omega, \qquad \frac{1}{H}\cdot \frac{1}{\zeta \beta_0}=b_0\sqrt{\Omega};\\
\frac{1}{H^2}\cdot\frac{1}{\beta_2} &= a_2 \Omega, \qquad \frac{1}{H}\cdot \frac{1}{2\eta \beta_2}=b_2\sqrt{\Omega}.
\end{split}
\end{align}
The parameter~$a_0$ is restricted from above by $a_0<3\gamma(2-\gamma)$. This limitation comes from the expression for the speed of bulk viscous perturbations derived in~\cite{Maartens1996}.
\par 
Introducing the quantities
\begin{equation}
\mathbf{X}_\Sigma = (\Sigma_+,\Sigma_-,\Sigma_{23}), \qquad
\mathbf{X}_\Pi	= (\Pi_+,\Pi_-,\Pi_{23})
\end{equation}
as vectors in~$\mathbb{R}^3$, we can write the equations in a more compact manner. The most general case we consider is a Bianchi type~IV cosmology with a dissipative fluid and vacuum energy having the form of a cosmological constant. In this case, the equations of motion are:
\begin{align}
\Sigma^\prime_+ &= (q-2)\Sigma_+ -2N^2+\Pi_+, \label{Eq:Eqs:EvolSigmaPlus} \\
\Sigma^\prime_- &= (q-2-2\sqrt{3}\Sigma_{23})\Sigma_-+2AN+\Pi_-,\\
\Sigma_{23}^\prime &= (q-2)\Sigma_{23} +2\sqrt{3}\Sigma_-^2-2\sqrt{3}N^2+\Pi_{23},\\
N^\prime &= (q+2\Sigma_++2\sqrt{3}\Sigma_{23})N, \\
A^\prime &= (q+2\Sigma_+)A.
\end{align}
The evolution equations for the fluid can be written as:
\begin{align}
\Omega^\prime &= (2q+2-3\gamma)\Omega-3\Pi-2(\mathbf{X}_\Sigma\cdot \mathbf{X}_\Pi),\\
\Omega_\Lambda^\prime &= 2(q+1)\Omega_\Lambda,\\
\Pi^\prime &= \left[-\frac{3}{2}+\frac{1}{\gamma}(q+1)-b_0\Omega^{1/2}+\frac{2\gamma-1}{2\gamma}\frac{\Omega^\prime}{\Omega} \right]\Pi-a_0\Omega, \label{Eq:Eqs:EvolBulkVisk} \\
\mathbf{X}_\Pi^\prime &= \left[-\frac{3}{2}+\frac{1}{\gamma}(q+1)-b_2\Omega^{1/2}+\frac{2\gamma-1}{2\gamma}\frac{\Omega^\prime}{\Omega} \right]\mathbf{X}_\Pi-2a_2\mathbf{X}_\Sigma, \label{Eq:Eqs:EvolShearVisk}
\end{align}
with the deceleration parameter~$q$ being given by
\begin{equation}
q=2\vert\mathbf{X}_\Sigma\vert^2+\left(\frac{3}{2}\gamma-1\right)\Omega+\frac{3}{2}\Pi-\Omega_\Lambda.
\end{equation}
In the absence of tilt and heat conduction, only two algebraic constraint survives:
\begin{align}
1 &= \vert\mathbf{X}_\Sigma\vert^2+N^2+A^2+\Omega+\Omega_\Lambda, \label{Eq:Eqs:HamiltonianConstraint} \\
0 &= \Sigma_+A+\Sigma_-N. \label{Eq:Eqs:RConstraint}
\end{align}
Equations (\ref{Eq:Eqs:EvolSigmaPlus})--(\ref{Eq:Eqs:RConstraint}) form a complete system which will be used for analytical and numerical investigations. Replacing equations~(\ref{Eq:Eqs:EvolBulkVisk}) and~(\ref{Eq:Eqs:EvolShearVisk}) respectively with
\begin{align}
\Pi^\prime &= \left[2(q+1)-b_0\Omega^{1/2}\right]\Pi-a_0\Omega, \label{Eq:Eqs:EvolBulkVisk-trunc} \\
\mathbf{X}_\Pi^\prime &= \left[2(q+1)-b_2\Omega^{1/2}\right]\mathbf{X}_\Pi-2a_2\mathbf{X}_\Sigma\label{Eq:Eqs:EvolShearVisk-trunc}
\end{align}
yields the system with the truncated IS~transport equations.

\par 
The equations for Bianchi type~V cosmologies are obtained by setting~$N=0$ in equations~(\ref{Eq:Eqs:EvolSigmaPlus})--(\ref{Eq:Eqs:RConstraint}). Note that it implies that for this model~$\Sigma_+=\Pi_+=0$. As for the case without vacuum energy, one simply assigns~$\Omega_\Lambda=0$.
\par 
Independently of whether one is considering the system with the full or truncated transport equations, the state vector describing a Bianchi type~IV cosmology containing a dissipative $\gamma$-fluid and vacuum energy can be written as
\begin{equation}
\mathbf{X}=(\mathbf{X}_\Sigma,N,A,\Omega,\Omega_\Lambda,\Pi,\mathbf{X}_\Pi)
\end{equation}
modulo the constraints~(\ref{Eq:Eqs:HamiltonianConstraint})--(\ref{Eq:Eqs:RConstraint}). The evolution takes place on a 9-dimensional subspace of~$\mathbb{R}^{11}$. The physical state space is therefore of dimension~9. Considering the simpler  Bianchi type~V model or the case without a cosmological constant reduces the dimension of the physical state space respectively by~2 and~1.

\par 
The stability of equilibrium points for the system~(\ref{Eq:Eqs:EvolSigmaPlus})--(\ref{Eq:Eqs:RConstraint}) is investigated by standard analytical methods, see e.\thinspace g.~\cite{Hervik2005, Shogin2014a}, and numerical simulations are then used to make a conjecture about the global attractor of this dynamical system. We also use numerical runs to determine the relative magnitude of the dissipative fluxes. In the current paper we shall omit technical details, focusing on the main results.

\section{The futures of Bianchi type IV and V universes without vacuum energy}
\label{Sec:AttractorsNoLambda}

\subsection{The full transport equations}
The future stability of equilibrium points of the full system with the non-truncated transport equations depends critically on the bulk viscous stresses. Based on the precise relation between the bulk viscosity parameters~$a_0$ and~$b_0$, the system will asymptotically tend to one of the following future attractors:
\begin{enumerate}

\item 

For
\begin{equation}
0<a_0<\frac{(3\gamma-2)^2}{6\gamma}
\end{equation}
the future asymptotic state is the Milne universe given by
\begin{align}
\begin{split}
\label{Eq:Results:Milne}
[\mathbf{X}_\Sigma,\mathbf{X}_\Pi,A,N] & = [0,0,1,0],\\
[\Omega,\Pi,q] &= [0,0,0].
\end{split}
\end{align}
The evolution of the universe is ultimately dominated by the three-space curvature, and the expansion is asymptotically uniform.

\item 

For 
\begin{equation}
\frac{(3\gamma-2)^2}{6\gamma}<a_0<\frac{(3\gamma-2)^2}{6\gamma}+\frac{3\gamma-2}{3} b_0
\end{equation}
we obtain the "transitionary" solution:
\begin{align}
\begin{split}
\label{Eq:Results:Transitionary}
[\mathbf{X}_\Sigma,\mathbf{X}_\Pi,A,N] & = [0,0,\bar{A},0],\\
[\Omega,\Pi,q] &= [\bar{\Omega},\bar{\Pi},0],
\end{split}
\end{align}
with
\begin{equation}
\bar{\Omega}={\left[\frac{a_0-\frac{(3\gamma-2)^2}{6\gamma}}{b_0(\gamma-\frac{2}{3})}\right]}^2, \qquad \bar{A}=\sqrt{1-\bar{\Omega}}, \qquad \bar{\Pi}=-\left( \gamma-\frac{2}{3} \right) \bar{\Omega}.
\end{equation}
The numerical simulations show that the geometric and viscous shear stresses decay extremely slowly in this case. The dynamics of the universe is anisotropic, but the system is slowly approaching isortopy at late times. The expansion is asymptotically uniform, the fluid, the shear stresses, and the three-space curvature all playing a significant role.

\item

For
\begin{equation}
a_0>\frac{(3\gamma-2)^2}{6\gamma}+\frac{3\gamma-2}{3} b_0
\end{equation}
the future attractor is
\begin{align}
\begin{split}
\label{Eq:Results:BVI}
[\mathbf{X}_\Sigma,\mathbf{X}_\Pi,A,N] & = [0,0,0,0],\\
[\Omega,\Pi,q] &= [1,\bar{\Pi},\bar{q}],
\end{split}
\end{align}
with
\begin{align}
\bar{\Pi} &= \frac{1}{3}\left[\gamma b_0-\sqrt{\gamma^2 b_0^2+6\gamma a_0}\right],\\
\bar{q} &= -1+\frac{1}{2}\left[3\gamma+\gamma b_0-\sqrt{\gamma^2 b_0^2+6\gamma a_0}\right].
\end{align}

In this case, the universe is dominated by the fluid. The future asymptotic state corresponds to the bulk viscous inflationary scenario. The universe is asymptotically flat, its expansion being accelerated; the acceleration is caused by the (asymptotically constant) negative bulk viscous term.

\item 
For the values on the border line between cases~1 and~2 above, where
\begin{equation}
a_0=\frac{(3\gamma-2)^2}{6\gamma},
\end{equation}
the future attractor is still the Milne universe given by~(\ref{Eq:Results:Milne}). However, as for the solutions tending to~(\ref{Eq:Results:Transitionary}), the dynamics of the universe is anisotropic. The system approaches isotropy very slowly at late times.

\item 
Finally, for the values on the border line between cases~2 and~3, where
\begin{equation}
a_0=\frac{(3\gamma-2)^2}{6\gamma}+\frac{3\gamma-2}{3} b_0,
\end{equation}
the future attractor is the boundary state between the transitionary solution~(\ref{Eq:Results:Transitionary}) and bulk viscous inflation~(\ref{Eq:Results:BVI}):
\begin{align}
\begin{split}
[\mathbf{X}_\Sigma,\mathbf{X}_\Pi,A,N] & = [0,0,0,0],\\
[\Omega,\Pi,q] &= [1,-(\gamma-2/3),0].
\end{split}
\end{align}
The universe is asymptotically dominated by the fluid with negative bulk viscous pressure, the space three-curvature tends to zero, the expansion being uniform. The future attractor is approached extremely slowly, and the model remains significantly anisotropic during its evolution.
\end{enumerate}

Our numerical simulations show that deviations from equilibrium, caused by bulk viscosity, tend to a finite constant value for all the solutions. At the same time, deviations, caused by shear viscous stresses, decay in cases~3 and~5 above, are effectively finite for the transitionary solutions~(case 2), and increase without bound for solutions tending to the Milne universe~(cases~1 and~4). In the latter case the fluid state is evolving infinitely far from equilibrium, where the IS~theory is not valid anyhow.

\subsection{The truncated transport equations}
In constrast to the system with the full transport equations, in the truncated case the future stability of the equilibrium points is not dictated by the bulk viscosity parameters only. Shear viscosity can cause instabilities that yield solutions running into a singularity at finite times. The nature of this singularity is that the energy density~$\Omega$ is able to cross the vacuum boundary and become negative; the transport equations thus break down. This is a strict mathematical property of the truncated equations which is absent in the full IS~theory.
\par 
In addition, the stable universes described by the system with the truncated transport equations have only three possible future attractors:

\begin{enumerate}

\item 

For
\begin{equation}
0<a_0<\frac{1}{3}(b_0-2)(3\gamma-2)
\end{equation}
the future asymptotic state is
\begin{align}
\begin{split}
\label{Eq:Results:Transitionary-trunc}
[\mathbf{X}_\Sigma,\mathbf{X}_\Pi,A,N] & = [0,0,\bar{A},0],\\
[\Omega,\Pi,q] &= [\bar{\Omega},\bar{\Pi},0],
\end{split}
\end{align}
with
\begin{equation}
\bar{\Omega}={\left[\frac{a_0+2(\gamma-\frac{2}{3})}{b_0(\gamma-\frac{2}{3})}\right]}^2, \qquad \bar{A}=\sqrt{1-\bar{\Omega}}, \qquad \bar{\Pi}=-\left( \gamma-\frac{2}{3} \right) \bar{\Omega}.
\end{equation}
Note that this corresponds to the "transitionary" solution of the non-truncated case. As for the full equations, the universe is significantly anisotropic during the evolution, but slowly approaches isotropy at late times.
\par 
Moreover, the following (significant) restriction should be imposed on the shear viscosity parameters to avoid instability of future equilibria and running into a singularity:
\begin{equation}
a_2>2-b_2\sqrt{\bar{\Omega}}.
\end{equation}

\item 

For
\begin{equation}
a_0>\frac{1}{3}(b_0-2)(3\gamma-2)
\end{equation}
the universe evolves towards
\begin{align}
\begin{split}
\label{Eq:Results:BVI-trunc}
[\mathbf{X}_\Sigma,\mathbf{X}_\Pi,A,N] & = [0,0,0,0],\\
[\Omega,\Pi,q] &= [1,\bar{\Pi},\bar{q}],
\end{split}
\end{align}
with
\begin{align}
\bar{\Pi} &= \frac{1}{6}\left[-(3\gamma-b_0)-\sqrt{(3\gamma-b_0)^2+12a_0}\right],\\
\bar{q} &= -1+\frac{1}{4}\left[3\gamma+b_0-\sqrt{(3\gamma-b_0)^2+12a_0}\right].
\end{align}
This solution corresponds to bulk viscous inflation. Note however, that despite certain similarities between solutions in the full and truncated cases, these solutions lead to different expressions for asymptotic values of the bulk viscous stress and the deceleration parameter, and thus to different late-time dynamics of the universe.
\par 
Additional restrictions on the shear viscosity parameters are also required in this case. They can be written as
\begin{align}
a_2 & > \frac{3}{4}(\gamma+\bar{\Pi}-2)\left[b_2-3(\gamma+\bar{\Pi})\right],\\
b_2 & > -3+\frac{9}{2}(\gamma+\bar{\Pi}).
\end{align}

\item 
For the border values between cases~1 and~2, given by the line
\begin{equation}
a_0=\frac{1}{3}(b_0-2)(3\gamma-2),
\end{equation}
the future attractor is
\begin{align}
\begin{split}
[\mathbf{X}_\Sigma,\mathbf{X}_\Pi,A,N] & = [0,0,0,0],\\
[\Omega,\Pi,q] &= [1,-(\gamma-2/3),0].
\end{split}
\end{align}
Similarly to the corresponding case of the full theory, this describes the boundary state between the "transitionary" solution~(\ref{Eq:Results:Transitionary-trunc}) and bulk viscous inflation~(\ref{Eq:Results:BVI-trunc}). The universe, as might be expected, isotropizes fairly slowly.
\par 
To avoid the instability of the given attractor, the following additional requirement must be met:
\begin{equation}
a_2>2-b_2.
\end{equation}

\end{enumerate}

We see that the future equilibria of the truncated IS~theory are qualitatively similar to those of the full theory, although exact expressions, decay rates, and stability conditions are essentially different. In addition, the Milne universe, which is one of the possible future attractors in the full theory, is unstable in the truncated theory.
\par 
The deviations from equilibrium are identical to those in the corresponding solutions in the full theory.

\section{The futures of Bianchi type IV and V universes with vacuum energy}
\label{Sec:AttractorsLambda}
\subsection{The full transport equations}

Similarly to the case with zero cosmological constant, the future of the universe is determined by bulk viscosity parameters alone. Depending on particular relation between~$a_0$ and~$b_0$, the solution tends to one of four possible future attractors.
\begin{enumerate}

\item 

For
\begin{equation}
0<a_0\leq\frac{3\gamma}{2}
\end{equation}
the universe ends up in de Sitter state:
\begin{align}
\begin{split}
[\mathbf{X}_\Sigma,\mathbf{X}_\Pi,A,N] & = [0,0,0,0],\\
[\Omega,\Omega_\Lambda,\Pi,q] &= [0,1,0,-1].
\end{split}
\end{align}

The expansion of the universe is accelerated, and the cosmological constant dominates the evolution; viscous terms do not play a significant role. Note that this is the future asymptotic state of Bianchi cosmologies with a {\it perfect} $\gamma$-fluid and a cosmological constant~\cite{Christiansen2008}. We also point at the fact that for the border value~$a_0=3\gamma/2$ the future attractor is the same, however the universe isotropizes at slower rates.

\item 
At
\begin{equation}
\frac{3\gamma}{2}<a_0<\frac{3\gamma}{2}+\gamma b_0
\end{equation}
the future attractor is
\begin{align}
\begin{split}
[\mathbf{X}_\Sigma,\mathbf{X}_\Pi,A,N] & = [0,0,0,0],\\
[\Omega,\Omega_\Lambda,\Pi,q] &= [\bar{\Omega},\bar{\Omega}_\Lambda,\bar{\Pi},-1],
\end{split}
\end{align}
with
\begin{equation}
\bar{\Omega}={\left[\frac{a_0-\frac{3\gamma}{2}}{\gamma b_0}\right]}^2, \qquad \bar{\Omega}_\Lambda=1-\bar{\Omega}, \qquad \bar{\Pi}=-\gamma \bar{\Omega}.
\end{equation}
This describes a "mixed" inflation with~$q\to-1$: both positive cosmological constant and fluid with negative bulk viscosity make a contribution to accelerating the expansion of universe.

\item 

For

\begin{equation}
a_0>\frac{3\gamma}{2}+\gamma b_0
\end{equation}
the future attractor is given by
\begin{align}
\begin{split}
[\mathbf{X}_\Sigma,\mathbf{X}_\Pi,A,N] & = [0,0,0,0],\\
[\Omega,\Omega_\Lambda,\Pi,q] &= [1,0,\bar{\Pi},\bar{q}],
\end{split}
\end{align}
with
\begin{align}
\bar{\Pi} &= \frac{1}{3}\left[\gamma b_0-\sqrt{\gamma^2 b_0^2+6\gamma a_0}\right],\\
\bar{q} &= -1+\frac{1}{2}\left[3\gamma+\gamma b_0-\sqrt{\gamma^2 b_0^2+6\gamma a_0}\right].
\end{align}

This corresponds to "pure" bulk viscous inflation, where the fluid is ultimately dominating over the cosmological constant. Note that for "large" viscosity parameters, the asymptotic futures of the models with and without a cosmological constant are exactly the same.

\item 
Finally, for the border values between cases~2 and~3 above, given by
\begin{equation}
a_0=\frac{3\gamma}{2}+\gamma b_0,
\end{equation}
the future attractor is
\begin{align}
\begin{split}
[\mathbf{X}_\Sigma,\mathbf{X}_\Pi,A,N] & = [0,0,0,0],\\
[\Omega,\Omega_\Lambda,\Pi,q] &= [\gamma^{-2},1-\gamma^{-2},-\gamma^{-1},-1],
\end{split}
\end{align}
which describes a special case of "mixed" inflation with~$q\to-1$.
\end{enumerate}

As in the case without a cosmological constant, the relative dissipative fluxe caused by bulk viscosity tends to a constant for all the solutions. However, the deviations from equilibrium caused by shear viscous stresses in presence of cosmological constant tend always to zero.
\subsection{The truncated transport equations}
Unlike the case without a cosmological constant, the presence of positive~$\Lambda$ does not allow the shear viscosity to affect the future of the universe. The stability of future attractors is determined by the bulk viscosity parameters. Namely, depending on the values of~$a_0$ and~$b_0$, the three future asymptotic states are possible:

\begin{enumerate}

\item
For
\begin{equation}
0<a_0<\gamma b_0
\end{equation}
the future asymptotic state corresponds to a "mixed" inflation:
\begin{align}
\begin{split}
[\mathbf{X}_\Sigma,\mathbf{X}_\Pi,A,N] & = [0,0,0,0],\\
[\Omega,\Omega_\Lambda,\Pi,q] &= [\bar{\Omega},\bar{\Omega}_\Lambda,\bar{\Pi},-1],
\end{split}
\end{align}
with
\begin{equation}
\bar{\Omega}=\frac{a_0^2}{\gamma^2 b_0^2}, \qquad \bar{\Omega}_\Lambda=1-\bar{\Omega}, \qquad \bar{\Pi}=-\gamma \bar{\Omega}.
\end{equation}

\item 
For
\begin{equation}
a_0>\gamma b_0
\end{equation}
the universe is undergoing "pure" bulk viscous inflation:
\begin{align}
\begin{split}
[\mathbf{X}_\Sigma,\mathbf{X}_\Pi,A,N] & = [0,0,0,0],\\
[\Omega,\Omega_\Lambda,\Pi,q] &= [1,0,\bar{\Pi},\bar{q}],
\end{split}
\end{align}
with
\begin{align}
\bar{\Pi} &= \frac{1}{6}\left[-(3\gamma-b_0)-\sqrt{(3\gamma-b_0)^2+12a_0}\right],\\
\bar{q} &= -1+\frac{1}{4}\left[3\gamma+b_0-\sqrt{(3\gamma-b_0)^2+12a_0}\right].
\end{align}

\item 
For the values on the border line between cases~1 and~2, given by
\begin{equation}
a_0=\gamma b_0,
\end{equation}
the universe approaches the state with
\begin{align}
\begin{split}
[\mathbf{X}_\Sigma,\mathbf{X}_\Pi,A,N] & = [0,0,0,0],\\
[\Omega,\Omega_\Lambda,\Pi,q] &= [1,0,-\gamma,-1],
\end{split}
\end{align}
which corresponds to the limiting case of bulk viscous inflation with~$q\to-1$.
\end{enumerate}

The consequences of using the truncated theory are somewhat similar to the case without a cosmological constant. The future equilibria of the truncated IS~theory, describing "pure" and "mixed" bulk viscous inflation, are qualitatively similar to the corresponding equilibria of the full theory. The quantitative difference is at the same time essential, and the stability conditions are substantially different. The de Sitter state, which is a possible future attractor in the full theory, is unstable in the truncated IS~theory.

\begin{figure}[ht!]
\begin{minipage}[h]{0.45\linewidth}
\includegraphics[width=1.0\linewidth]{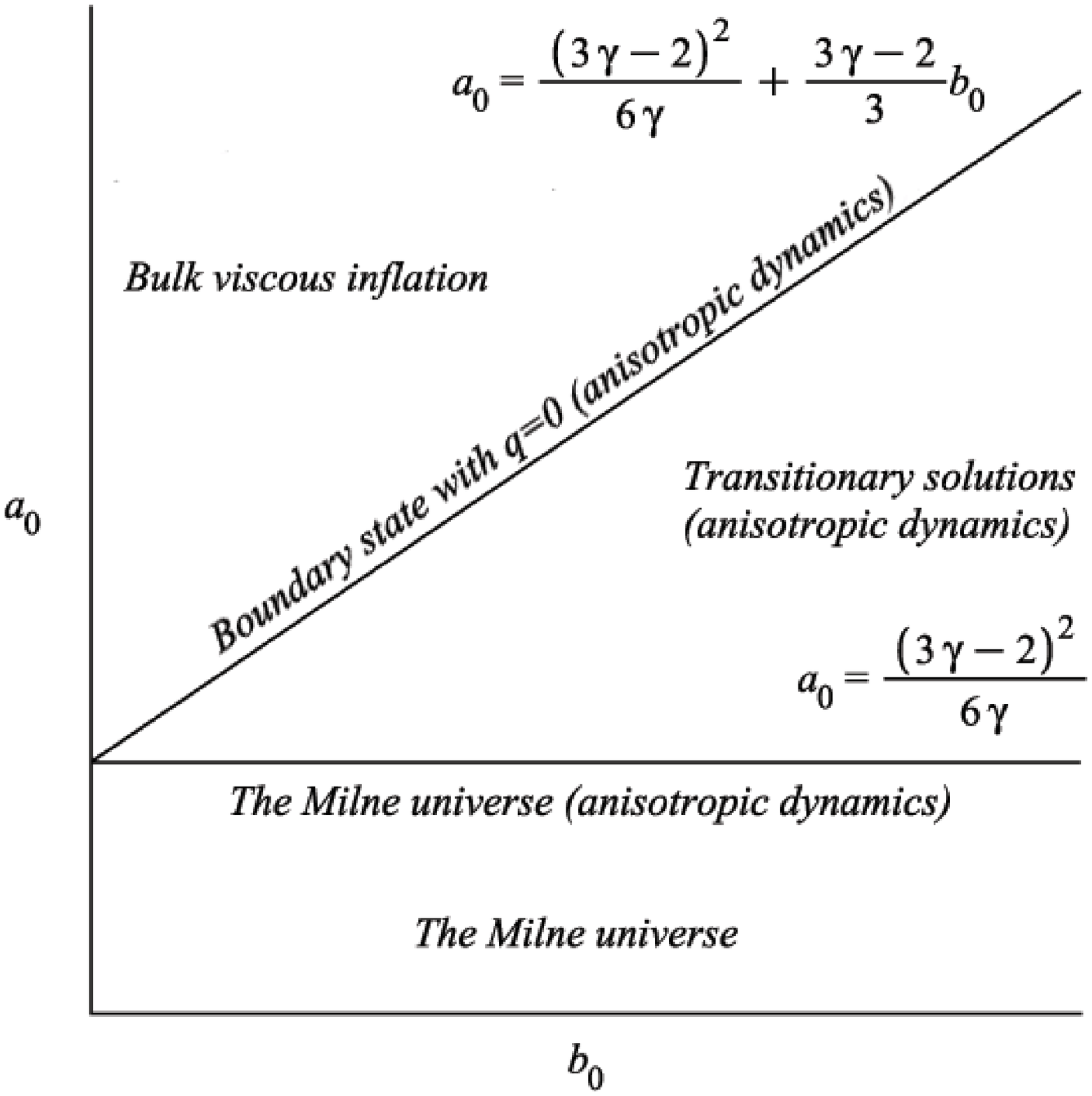}
\end{minipage}
\begin{minipage}[h]{0.45\linewidth}
\includegraphics[width=1.0\linewidth]{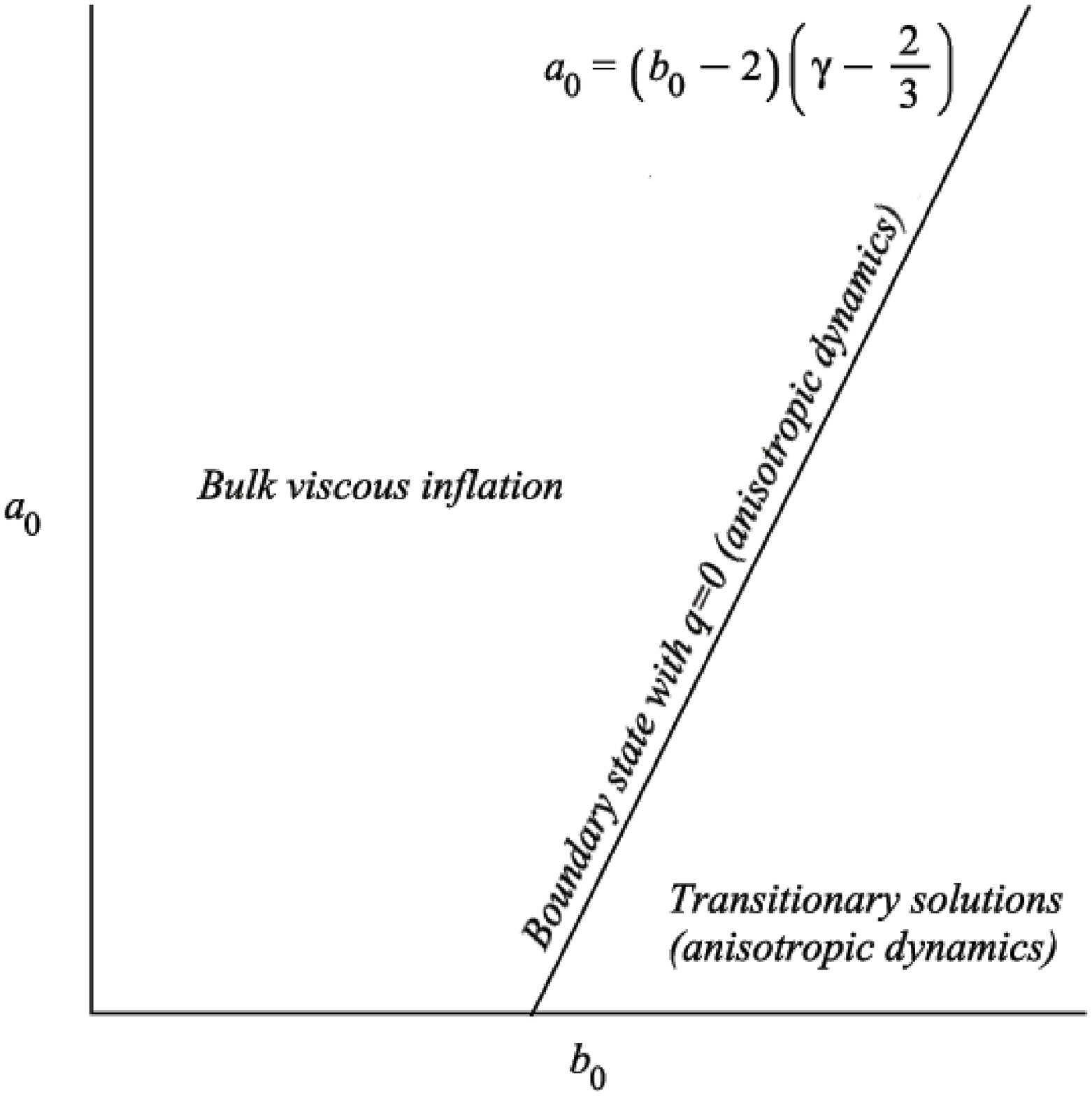}
\end{minipage}
\caption{The asymptotic futures of Bianchi type~IV and~V cosmological models without vacuum energy as functions of the bulk viscosity parameters, as predicted by the full (left) and truncated (right) IS~theories. Only non-singular solutions of the truncated theory are considered. Note that the state corresponding to the Milne universe is unstable in the truncated theory}
\label{Fig:NoLambda}
\end{figure}

\begin{figure}[ht!]
\begin{minipage}[h]{0.45\linewidth}
\includegraphics[width=1.0\linewidth]{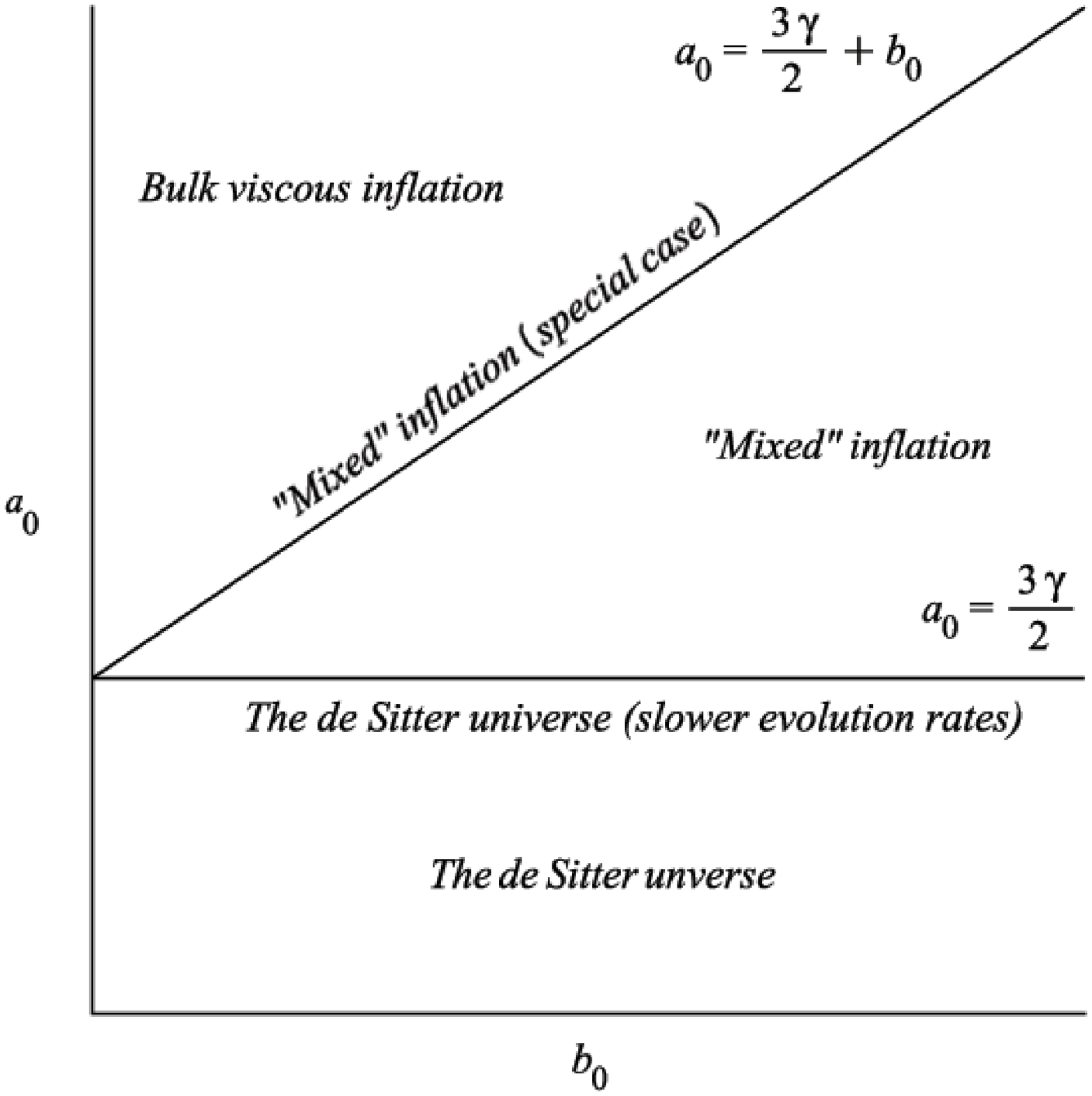}
\end{minipage}
\begin{minipage}[h]{0.45\linewidth}
\includegraphics[width=1.0\linewidth]{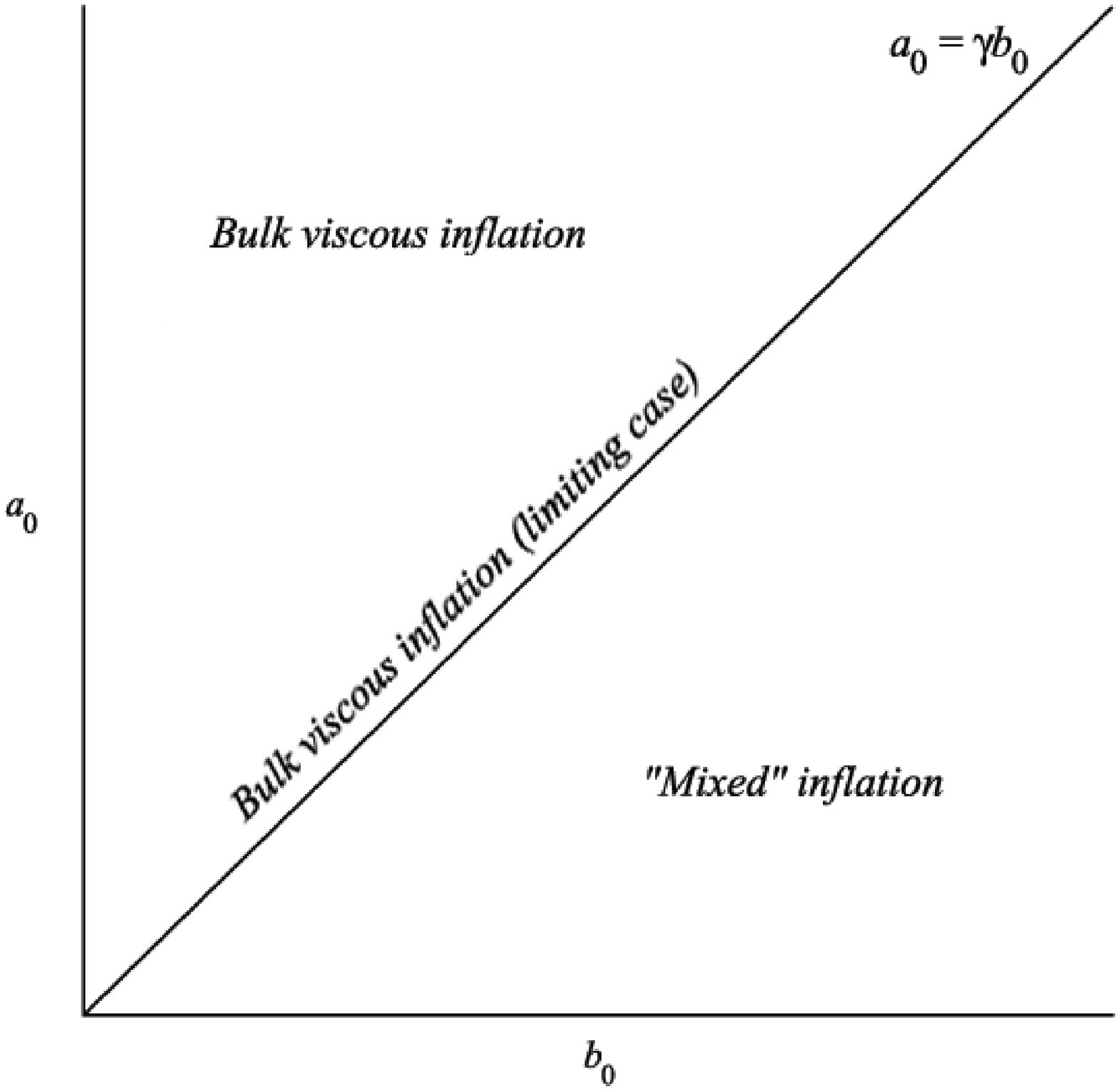}
\end{minipage}
\caption{The asymptotic futures of Bianchi type~IV and~V cosmologies with vacuum energy as functions of the bulk viscosity parameters, as predicted by the full (left) and truncated (right) IS~theories. Only non-singular solutions of the truncated theory are considered. Note that the de Sitter state becomes unstable in the truncated theory}
\label{Fig:Lambda}
\end{figure}

\section{Conclusion}
\label{Sec:Conslusion}

We have investigated the future dynamics, focusing on the equilibrum states and their stability conditions, for Bianchi type~IV and~V cosmologies with space filled with a dissipative~$\gamma$-fluid and, possibly, vacuum energy, which is modelled by a positive cosmological constant. We have considered the full transport equations of the IS~theory as well as the widely used truncated version of these equations. The future equilibrium states of the models as functions of the bulk viscosity parameters are presented in Figures~\ref{Fig:NoLambda} and~\ref{Fig:Lambda}. Multiple numerical runs allow us to make a conjecture that these states represent the global attractor for the system.
\par 
We have found that two bifurcations are present in the full IS~theory, while the truncated theory has only one. As a result, the Milne universe, which is one of possible future asymptotic states for the models without a cosmological constant in the full theory, is unstable in the truncated theory. Correspondingly, for the models with a cosmological constant, the de Sitter universe becomes unstable in the truncated theory. The other future equilibria of both theories are qualitatively similar, but the quantitative expressions, the stability conditions, and, therefore, the dynamics of the models are substantially different.
\par 
We have also found that the truncated transport equations possess mathematical properties which may have unphysical consequences. Shear viscosity can become a source of instability of the future equilibria, and cause universe models without a cosmological constant to run into a singularity. This phenomenon is not present in the full theory. In order to avoid singular solutions of the truncated theory, severe restrictions must be imposed on the shear viscosity parameters.
\par 
In conclusion, great care must be taken if the truncated IS~theory is employed in a cosmological setting, in particular in the case of anisotropic cosmological models. It is true that the fluid description provided by the IS~theories is multiparametric, the future equilibrium states and their stability being dependent of the chosen equation of state and physical models of the transport coefficients. Still, this does not change our conclusion that the truncation of the IS~transport equations in general leads to very different asymptotic states and dynamics of the solutions.
\par 
According to our numerical results, in all the solutions obtained, the fluid is strongly out of equilibrium, which implies that the underlying assumption of the IS~theory ceases to be valid. The deviations from the equilibrium, caused by bulk viscosity, are found to be always finite; the problem might therefore be solved by a modification of the~IS transport equations or of the equation of state, using the effective pressure instead of the local equilibrium value, at least in some cases. However, the relative dissipative fluxes, caused by shear viscosity, can either decay, be finite or increase without bound. In the latter case, the IS~theory is strictly inapplicable. This underlines the importance of a consistent theory of nonlinear thermodynamics and provides a possible direction for further investigations. 
\bibliography{Bib/New}
\end{document}